# Spectral Scaling in Complex Networks

Ernesto Estrada

*Complex Systems Research Group, X-ray Unit, RIAIDT, Edificio CACTUS, University of Santiago de Compostela, 15782 Santiago de Compostela, Spain.*

A complex network is said to show *topological isotropy* if the topological structure around a particular node looks the same in all directions of the whole network. *Topologically anisotropic networks* are those where the local neighborhood around a node is not reproduced at large scale for the whole network. The existence of topological isotropy is investigated by the existence of a power-law scaling between a local and a global topological characteristic of complex networks obtained from graph spectra. We investigate this structural characteristic of complex networks and its consequences for 32 real-world networks representing informational, technological, biological, social and ecological systems.

The study of complex networks represents a unifying language to describe systems in disparate real-world contexts ranging from biological to technological systems. A plethora of works have been dedicated toward understanding the topological and dynamical properties of these networks [1-7]. Two classes of models, the so-called *small-world graphs* [8, 9] and *scale-freeness networks* [10, 11], have been developed to capture some of the general properties of real-world networks. Most of these analysis have been limited to static structural characteristics or statistical parameters which are only loosely connected to structural properties, such as degree sequences, average shortest paths, cliquishness, etc. However, it is known that global measures of the network properties are well characterized by spectral methods [12, 13]. Here, we will use spectral graph theory to study the global topological structure of real-world complex networks. We start by considering a hypothetical physical process based on the emission of a signal from a particular node, which is transmitted from node to node by the links connecting them visiting $l$ (not necessarily different) edges of the network. The trajectory of this signal is described topologically by the number of *walks* in the network. A walk of length $l$ is any sequence of (not necessarily) different vertices $v_1, v_2, \cdots, v_l, v_{l+1}$ such that for each $i = 1, 2, \cdots, l$ there is an edge from $v_i$ to $v_{i+1}$. In the cases in which this signal returns to the source at the end of the trajectory it completes a closed walk in the network. A *closed walk* (CW) of length $l$ is a walk in which $v_{l+1} = v_1$.

We will consider that the topological structure of the network can be mapped from the information provided by the returned signal. In the simplest case — a regular lattice — it will be enough that the signal goes only to a close neighbourhood of the source in order to map the whole topological structure of the network. This is a consequence of the *topological isotropy* of this network. By topological isotropy we will understand that the topological structure around a particular node looks the same in all directions of the network. On the contrary, it could be possible that the information obtained by the signal on the neighbourhood of the source is not reproduced at the large scale for the whole network. In this kind of networks, which will be called *topologically anisotropic*, the signal needs to visit all the nodes to obtain a global topological picture of the network. Thus, it is necessary to explore in which extension real-world complex network are topologically isotropic or not in order to understand their general topological structure.

From the topological point of view this problem can be stated as follows. Let $M_1$ and $M_2$ be topological measures characterizing the local neighbourhood around a node and the global topological structure of a network, respectively. Then, isotropic networks will be characterized by a scaling relationship between $M_1$ and $M_2$ indicating that the local structure around the neighbourhood of a node is similar to its global neighbourhood. The lack of this scaling indicates that the network is anisotropic and "what you see locally around a node is not what you get globally for the network".

As a local measure $M_1$ we propose to use the *subgraph centrality* of a node $i$ in the network, $SC(i)$ [14]. This concept is based on the sum of all CWs of different lengths in the network starting (and ending) at node $i$. The number of CWs of length $l$ starting at vertex $i$ can be obtained from the graph spectrum as follows [14]:

$$\mu_l(i) = \sum_{j=1}^{N} [v_j(i)]^2 (\lambda_j)^l$$

where $v_j(i)$ is the $i$th component of the $j$th eigenvector of the adjacency matrix **A** and $\lambda_j$ is the corresponding $j$th eigenvalue [15]. Each of these CWs of length $l$ describes a subgraph of $m \leq l$ edges, which relates the number of CWs to the number of subgraphs in the network.

This is traduced in our model into the topological information obtained by the signal which is travelling through the network. The local character of this measure is provided by the fact that the CWs of length $l$ are divided by $l!$, which means that CWs of shorter lengths receive more weight in the sum than larger CWs. Consequently, the smaller subgraphs including node $i$, which describes the local neighbourhood of this node, receive more weight in $SC(i)$ than the larger subgraphs, which are describing a more global structure of the network. Using spectral graph theory we have shown that $SC(i)$ can be obtained as follows [14]:

$$SC(i) = \sum_{l=0}^{\infty} \frac{\mu_l(i)}{l!} = \sum_{j=1}^{N} [v_j(i)]^2 e^{\lambda_j}$$

It should be noted that $SC(i)$ counts all CWs in the network, which can be of even or odd length. CWs of even length might be trivial on moving back and forth in acyclic subgraphs, i.e., those that do not contain cycles, while odd CWs are always non-trivial, i.e., they do not contain contributions from acyclic subgraphs. Thus, it should be more appropriate to use only the sum of odd CWs as our $M_1$ measure. It is easy to show that:

$$SC(i) = \sum_{j=1}^{N} [v_j(i)]^2 \cosh(\lambda_j) + \sum_{j=1}^{N} [v_j(i)]^2 \sinh(\lambda_j) = SC_{even}(i) + SC_{odd}(i)$$

which means that the term $SC_{odd}(i)$ only accounts for subgraphs containing at least one odd cycle. In this way $SC_{odd}(i)$ can be considered as a local property of order in networks that characterise the odd-cyclic wiring of a typical neighbourhood.

As the global topological parameter $M_2$ we propose to use another graph spectral measure: the *main eigenvector* of the adjacency matrix. This parameter is also known as the *eigenvector centrality* [16] of the network. That is, let $G$ be a connected network and let $\lambda_1$ and $\mathbf{e}_1$, respectively, be the principal eigenvalue of $\mathbf{A}$ and the eigenvector corresponding to it, i.e., the *principal eigenvector* of $\mathbf{A}$ [15]. The global topological nature of the eigenvector centrality and its analogy with a signal travelling through all nodes of the network is provided by the following interpretation. The eigenvector centrality, $e_1(i)$, represents the probability that the number of walks of length $l$ starting at a vertex $i$ chosen at random be equal to $N_l$ for $l \to \infty$. This interpretation is straightforward from the following theorem of spectral graph theory [17]:

**Theorem:** Let $N_l(i)$ be the number of walks of length $l$ starting at node $i$ of a non-bipartite connected graph with $N$ vertices. Let $p_l(i) = N_l(i) / \sum_{j=1}^{N} N_l(j)$. Then $\lim_{l \to \infty} \mathbf{p}_l = \mathbf{e}_1$ where $\mathbf{p}_l = [p_l(1) \ p_l(2) \ \cdots \ p_l(N)]$ and $\mathbf{e}_1 = [e_1(1) \ e_1(2) \ \cdots \ e_1(N)]$ is the eigenvector corresponding to the index of $G$.

It is obvious that some of the walks counted by $N_l$ can go and return infinitely among the nodes of small subgraphs, such as a single link, connected triples, triangles, etc. However, these walks can also travel between the nodes of larger subgraphs, including travelling through the nodes of the whole network, all with the same weight. Consequently, $e_1(i)$ contains information about the short- and long-range topological structure of the neighbourhood of a node. Thus, $e_1(i)$ represents a characterization of the global structure of a network seen from node $i$ by mean of all walks (not necessarily closed) starting at this node and travelling through the whole network.

The existence of a scaling relationship of the type considered here is explored by mean of the local, $SC_{odd}(i)$, and global, $e_1(i)$, spectral characterization of the network structure.

This relationships has been studied for 32 complex networks of different types and sizes. These datasets include two semantic networks based on Roget's Thesaurus of English (Roget) and the Online Dictionary of Library and Information Science (ODLIS); five social networks that include a scientific collaboration network in the field of computational geometry, inmates in prison, injecting drug users (IDUs), Zachary karate club and college students on a course about leadership; four bibliographic citation (information) networks, one consisting of papers published in the *Proceedings of Graph Drawing* in the period 1994–2000, papers published in the field of "Network Centrality", papers citing Milgram's 1967 *Psychology Today* paper or using "Small World" in the title and papers published or citing articles from *Scientometrics* for the period 1978–2000; the airport transportation network in the US in 1997; the Internet at the autonomous systems (AS) level from September 1997 and April 1998; three networks of secondary-structure elements adjacency for large proteins; two protein–protein interaction networks (PINs), one for *Saccharomyces cerevisiae* (yeast) and the other for the bacterium *Helicobacter pylori*; three transcription interaction networks concerning *E. coli*, yeast and sea urchins; a neural network in *C. elegans*; 9 food webs representing a wide range of species numbers, linkage densities, taxa, and habitat types. These webs are Grassland, Scotch Broom, Ythan Estuary with parasites, El Verde Rainforest, St. Marks Seagrass, St. Martin Island, Little Rock Lake, Bridge Brooks Lake, Coachella Valley and Skipwith Pond.

Remarkably, in several of these complex networks we found a universal power-law scaling of the form $e_1(i) \propto [SC_{odd}(i)]^{\eta}$ with a robust exponent of 0.5 (see Figure 1).

This means that $e_1(i)$ scales to the infinite sum of non-trivial closed walks of length $(2l+1)$ divided by $(2l+1)!$, which start (and end) at node *i*. Thus, in these complex networks the probability of randomly finding a node that is the starting point for $N_l$ walks of length *l* (for $l \rightarrow \infty$) is proportional to the weighted sum of odd-cyclic subgraphs that contain node *i*. The graph spectral scaling observed here for complex networks suggests that the local structure around the neighbourhood of a node is similar to its global neighbourhood. We then say that "what you see locally is what you get globally" around a node of these networks. In other words, they show *topological structural isotropy*.

A different picture is observed if we consider a manifestly clustered network. As expected in these cases a lack of correlation appears between the local and global topological structures around nodes of the network. These networks show *topological structural anisotropy* and the power-law correlation between $e_1(i)$ and $SC_{odd}(i)$ is not observed for the whole network. This anisotropy is clearly observed for the Grassland food web, the social network of IDUs, the metabolic networks, and the secondary structure interaction networks in proteins, which are highly clustered networks. These plots do not consist of single straight lines but of several clusters of points grouping together some of the nodes (Figure 2).

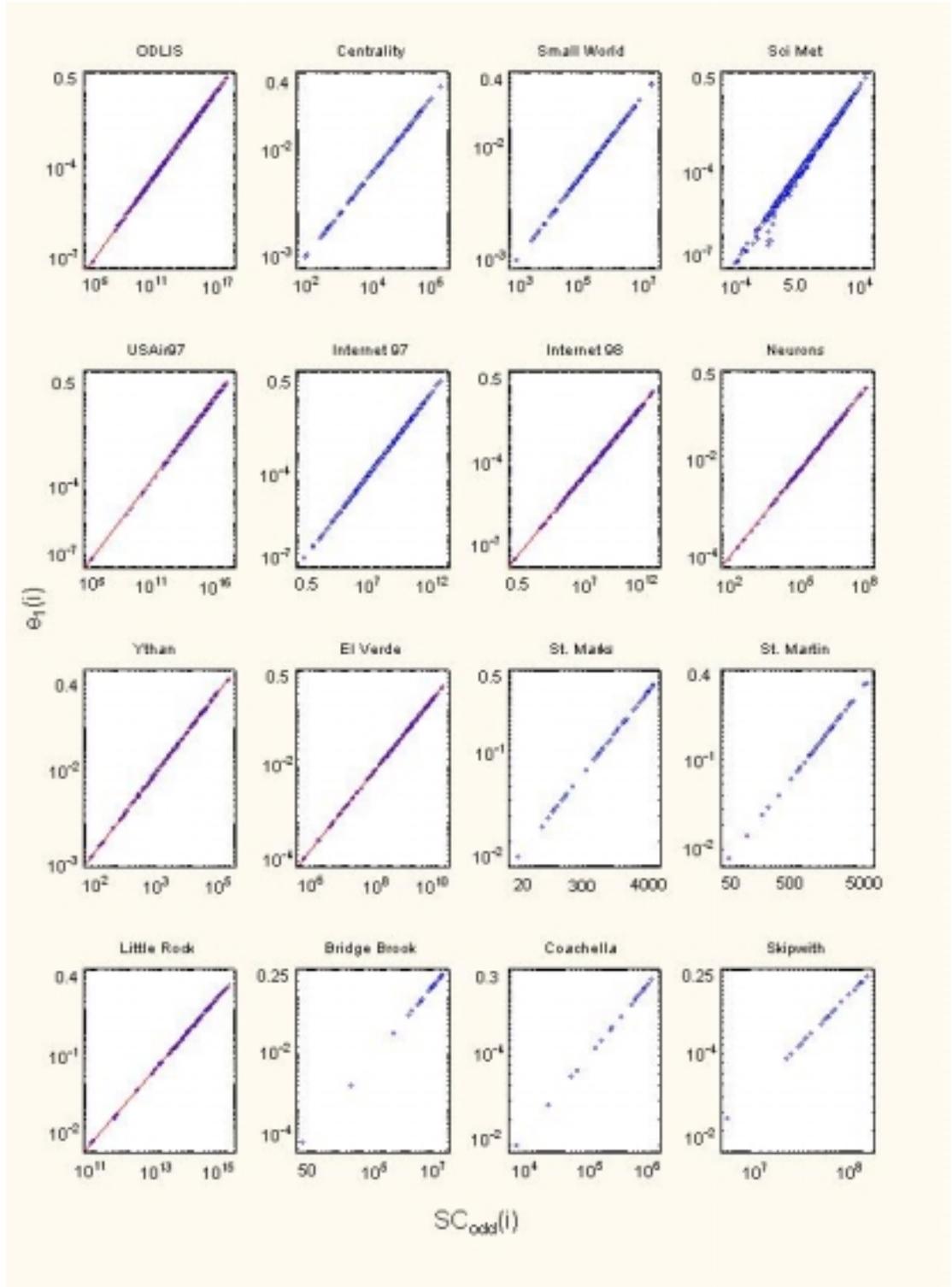

**FIG. 1.** Plot of the eigenvector centrality of node $i$, $e_1(i)$, versus the odd-cyclic subgraph centrality, $SC_{odd}(i)$, in a log-log plot for isotropic networks showing a scaling of the form: $e_1(i) \propto [SC_{odd}(i)]^\eta$. The universal scaling observed means that the global topological structure around a node is proportional to the local topology close to this node (see text).

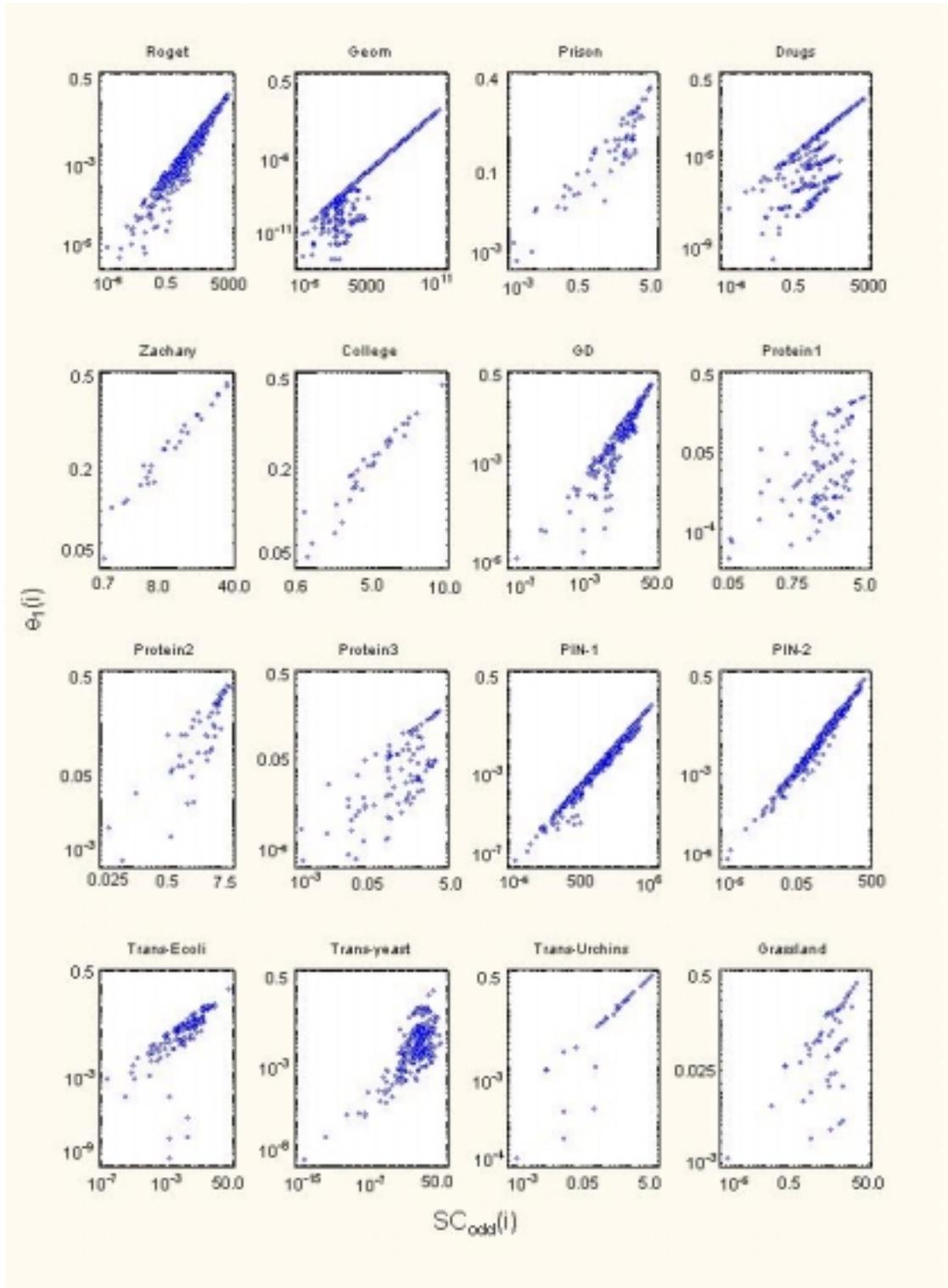

**FIG. 2.** Plot of the eigenvector centrality of node $i$, $e_1(i)$, versus the odd-cyclic subgraph centrality, $SC_{odd}(i)$, in a log-log plot for anisotropic networks. The lack of linear correlation means that what you see locally is not what you get globally around a node due to the high modularity of these networks (see text).

For instance, it is well known that proteins are built up from domains in a modular architecture [18]. Thus, the secondary structure elements manifest a higher number of within-domain than between-domain interactions, a situation that is translated into a highly clustered structure for the network. It has also been shown that most metabolic networks are organised into many small clusters of highly interconnected nodes [19]. These modules are combined in a hierarchical way into larger units that are less cohesive. In general, this kind of hierarchical modularity is present in all cellular networks that have been investigated to date [4]. In fact, a certain level of modularity is also observed in the PINs studied here, but in a less marked way than that observed for metabolic or secondary structure interaction networks. Manifest modularity is also observed in the social network of IDUs and in the Grassland food web (Figure 2), where several highly cohesive groups are distinguished as clusters with small interactions between each other (graphic of the network not shown). All social networks analysed are topologically anisotropic and show some social communities within them (Figure 2), which is probably a natural characteristic of social organization. However, random graphs with the same degree distribution that anisotropic networks are not able to reproduce their modularity and are manifestly isotropic (Figure 3). In contrast, most food webs (the only exception is Grassland) are topologically isotropic and represent ecosystems that are not composed of separate communities (FIG. 1) [20]. The relation between the current spectral scaling and the modularity of complex networks measured by statistical properties based on the average clustering coefficient [19] need to be studied in the future. However, we remark that the current scaling represents an essential topological (non-statistical) property of complex networks revealing organizational principles of their construction. As a simple example we can consider the existence of two types of networks with isotropic and anisotropic characteristics, respectively, which have clustering coefficient equal to zero as a consequence of the lack of triangles in the network. It is obvious that the average clustering coefficient does not distinguish between these two types of global topologies despite they are different by definition.

To conclude, in this work we have introduced a theoretical methodology for exploring the global topological structure of complex networks using spectral graph theory. This methodology consists in characterizing the local neighbourhood around a node and the global structure of the network, respectively, by using two spectral graph measures. A power-law scaling relationships between these two spectral characteristics of networks permits to classify all complex networks in two groups. *Topologically isotropic networks* are those where the topological structure around a particular node looks the same in all directions of the network, while in *topologically anisotropic networks* the local neighborhood around a node is not reproduced at large scale for the whole network. A direct consequence of this power-law scaling is that topologically isotropic networks are expected to be less sensitive to the absence of complete information about the structure of the network than topologically anisotropic networks. In isotropic networks one can study the global topological structure around a node by knowing its local topological structure. However, this situation is not possible in highly clustered networks for which local and global structures do not scale, which has great implications for studying different physical processes on such networks.

The author thanks J. A. Dunne, R. Milo, U. Alon, J. Moody, V. Batagelj and D. J. Watts for providing datasets. Useful comments made by A. Vázquez, S. Iztkovitz and J. A. Rodríguez-Velázquez are also acknowledged, as are clarifying discussions with S. N. Dorogovtsev and S. Valverde. This work was partially supported by the "Ramón y Cajal" program, Spain.

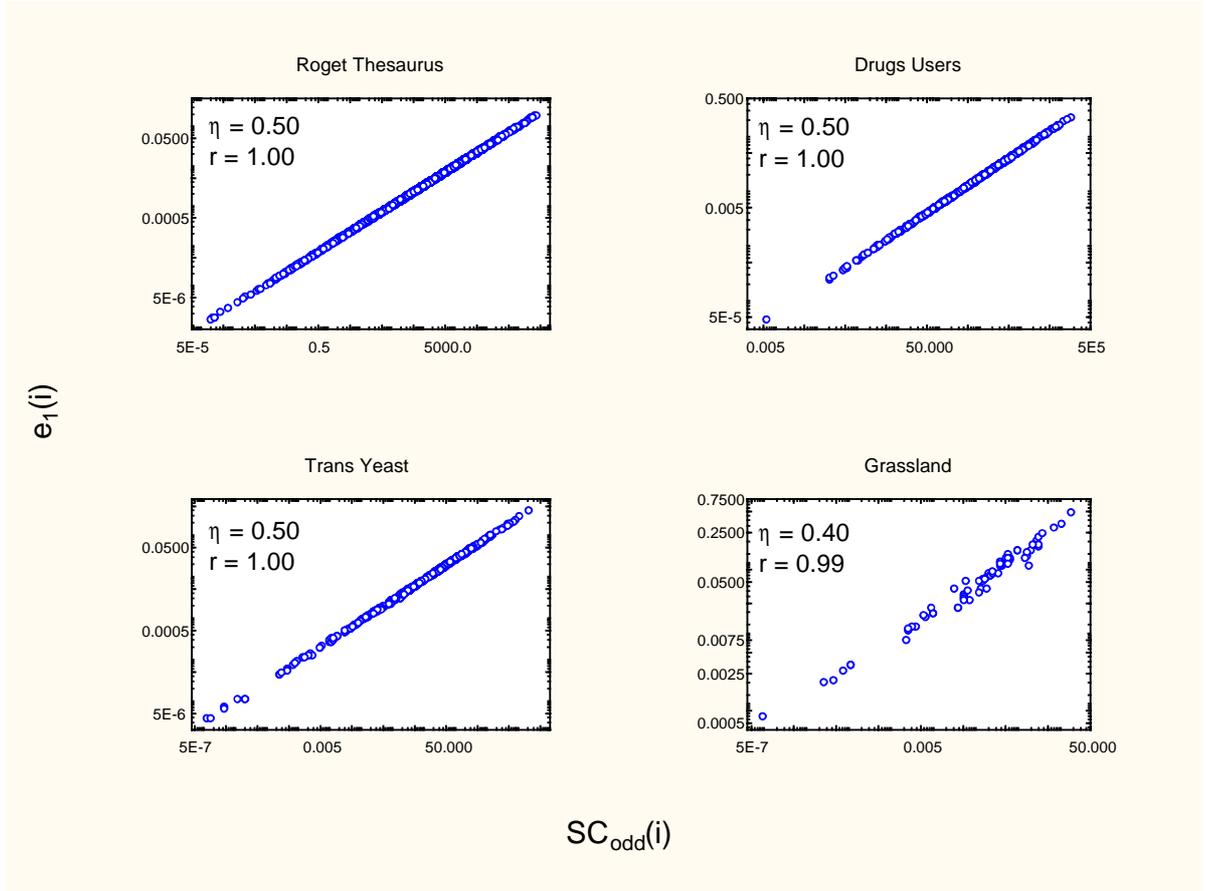

**Fig. 3.** Log-Log plots of the eigenvector centrality of node $i$, $e_1(i)$, versus the odd-cyclic subgraph centrality, $SC_{odd}(i)$, for random networks with the same degree distribution as the real-world networks representing Roget thesaurus of English (information), IDUs (social), metabolic network of yeast (biological) and Grassland (food web). The power-law coefficients $\eta$ and the correlation coefficient of the regression model, $r$, are also given. The straight line correlations indicate the existence of the scaling of global topological structure as a power-law of the local cliquishness around a node of the network. The original networks, however, show plots with several clusters of points indicating that these networks are highly modular (see Fig. 2).